\documentclass[epj,referee]{svjour}
\usepackage{graphics}
\usepackage{amsmath}
\usepackage{graphicx}
\usepackage{bm}
\usepackage{latexsym}
\usepackage{amssymb}
\usepackage{epsfig}
\usepackage{indentfirst}
\begin{document}
\title{Eradicating Computer Viruses on Networks}
\author{Jinyu Huang\inst{1} 
\thanks{\emph{e-mail:} jhuang14@iit.edu}%
}                     
%
%
\institute{$^1$Department of Applied Mathematics, Illinois Institute
of Technology, Chicago, Illinois 60616, USA}
%
%
\abstract{ Spread of computer viruses can be modeled as the SIS
(susceptible-infected-susceptible) epidemic propagation. We show
that in order to ensure the random immunization or the targeted
immunization effectively prevent computer viruses propagation on
homogeneous networks, we should install antivirus programs in every
computer node and frequently update those programs. This may produce
large work and cost to install and update antivirus programs. Then
we propose a new policy called ``network monitors" to tackle this
problem. In this policy, we only install and update antivirus
programs for small number of computer nodes, namely the ``network
monitors". Further, the ``network monitors" can monitor their
neighboring nodes' behavior. This mechanism incur relative small
cost to install and update antivirus programs.We also indicate that
the policy of the ``network monitors" is efficient to protect the
network's safety. Numerical simulations confirm our analysis.
\PACS{
      {89.75.Hc}{Networks and genealogical trees}   \and
      {05.70.Jk}{Critical point phenomena}   \and
      {64.60.ah}{Percolation}
     } 
} 
\maketitle
\section{Introduction}
\label{sec:1} One of key issues in the field of epidemiology is to
find effective strategies to prevent epidemic outbreaks. Researchers
widely studied immunization strategies in the SIS
(susceptible/infective/susceptible) model and the SIR
(susceptible/infective/removed) model. On some homogeneous networks,
random immunization can be applied to prevent epidemic propagation.
It was confirmed by many studies \cite{1,2,3}. However, on
heterogeneous networks just like scale-free (SF) networks, random
immunization mechanics is not particularly effective and we must
introduce other strategies to fight against epidemic. In order to
solve this problem, some researchers introduced optimal immunization
strategies such as the targeted immunization strategy and the
acquaintance immunization strategy. Relevant studies showed that
these immunization strategies are efficient on SF networks
\cite{2,3,4}.

For the computer network, We often use the SIS model to describe
computer viruses spreading on networks. In addition, relevant
immunization strategies can be applied to this model to prevent
computer viruses spreading. In this case, the situation that some
computer recovers and returns to the susceptible state can be
regarded as the fact that there is antivirus software in a computer
node and the computer can be ``cured" by this software if it is
infected by computer viruses. However, if we want relevant
immunization strategies such as the random immunization and the
targeted immunization to be effective on homogeneous networks, we
must install antivirus programs in each computer nodes and
frequently update those antivirus programs. Otherwise, random
immunization and targeted immunization may fail to prevent computer
viruses propagation. As a result, it may produces large costs to
install and update antivirus programs for each computer and
sometimes installing and frequently updating every computer's
antivirus programs is even impossible. For instance, the university
cannot ensure every student's computer or notebook in the campus
network possesses antivirus software. Consequently, it is meaningful
to find other efficient schemes that incurs relatively low costs to
install and update antivirus programs. In this paper, we concentrate
on this issue.

The paper is organized into following sections. In section
\ref{sec:2}, we show that if we cannot install and frequently update
antivirus programs for every computer on the homogeneous networks,
the random immunization strategy and the targeted immunization
strategy may fail to prevent virus propagation. Then, we introduce
``network monitor" strategy to protect computer network's safety and
analyze its efficiency on homogeneous networks. In section
\ref{sec:3}, we concentrate on the issue of the effectiveness of the
``network monitor" strategy on SF networks. Finally, we give
discussion and conclusions.
\section{Eradicating epidemics on homogeneous networks}
\label{sec:2} There are two classic ways to deal with problems of
the epidemic propagation. One way is based on the percolation theory
and techniques of generating functions \cite{5,6,7,8}. In fact, in
the study of the resilience of the network \cite{9,10,11,12} this
method was also widely used. The other way is based on mean-field
theory. In this paper, we mainly use mean-field theory for analysis.

First, we show that random immunization is not effective to prevent
computer viruses propagation if we cannot install and update
antivirus programs for each computer nodes in the network. In the
SIS model, computer nodes are classified as susceptible nodes and
infective nodes. Each susceptible node is infected with rate
$\lambda$ provided that it is connected with one or more infective
nodes. Without loss of generality, we also assume that infective
nodes are cured and become susceptible nodes with rate 1. Here,
infective nodes are ``cured" by antivirus programs. For simplicity,
we concentrate on the WS network \cite{13} with rewire probability
$1$, which has similar degree distribution with the ER random graph
model \cite{14} and can be viewed as a homogeneous network. Then the
differential equation of the SIS model is as follows:
\begin{equation}
\frac{\mathrm{d}\rho(t)}{\mathrm{d}t}=\lambda\langle k \rangle
\rho(t)[1-\rho(t)]-\rho(t)
\end{equation}
Here, $\rho(t)$ represents the density of infected nodes, $\langle k
\rangle$ is the average degree of a homogeneous network and
$\lambda$ is the spreading rate. Suppose we can only install and
update antivirus programs for $\alpha$ ($0<\alpha<<1$) computer
nodes ($\alpha$ is the fraction of computer nodes). Then $1-\alpha$
computer nodes cannot recover when they become infective. In this
case, Eq.\,(1) is not appropriate to describe propagation behavior.
The reason is as follows: the density of $\rho(t)$ is somewhat
random density, then they may belong to all $1-\alpha$ computer
nodes, which means that they cannot recover again. As a result, the
second item on the right of Eq.\,(1) is invalid. If $1-\alpha$
vertices on the network cannot recover after they become infective,
it is better to use SI (susceptible/infective) model to describe
virus propagation behavior among those $1-\alpha$ vertices. Without
loss of generality, we assume those $\alpha$ nodes are randomly
chosen from the WS network or the random network. Then the
subnetwork with $1-\alpha$ nodes is also random network. Thus, we
can just consider the effectiveness of random immunization with the
SI model on the WS network with rewire probability $1$. The
differential equation of the SI model is as follows:
\begin{equation}
\frac{\mathrm{d}\rho(t)}{\mathrm{d}t}=\lambda\langle k \rangle
\rho(t)[1-\rho(t)]
\end{equation}
If we apply random immunization, we can simply replace $\lambda$ by
$\lambda(1-g)$ where $g$ represents the immunity. Then the following
differential equation can be obtained:
\begin{equation}
\frac{\mathrm{d}\rho(t)}{\mathrm{d}t}=\lambda(1-g)\langle k \rangle
\rho(t)[1-\rho(t)]
\end{equation}
After imposing the stationary condition
$\frac{\mathrm{d}\rho(t)}{\mathrm{d}t}=0$, we have:
\begin{equation}
\lambda(1-g)\langle k \rangle \rho(t)[1-\rho(t)]=0
\end{equation}
The critical immunization can be obtained by letting Eq.\,(4) only
has one solution $\rho(t)=0$. It can be fulfilled by forcing
$\Delta=0$ where $\Delta$ represents the discriminant of Eq.\,(4).
So the critical immunization $g_{c}=1$ when $\lambda\neq0$. It shows
that random immunization is not effective.

\begin{figure}
  \centering
  \includegraphics[width=8.3cm]{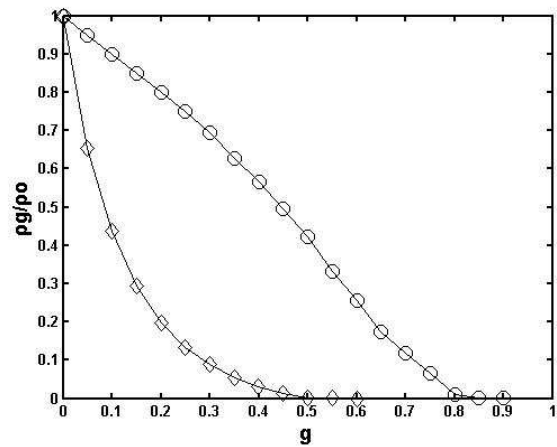}
  \centerline{(a)}\\
  \includegraphics[width=8.3cm]{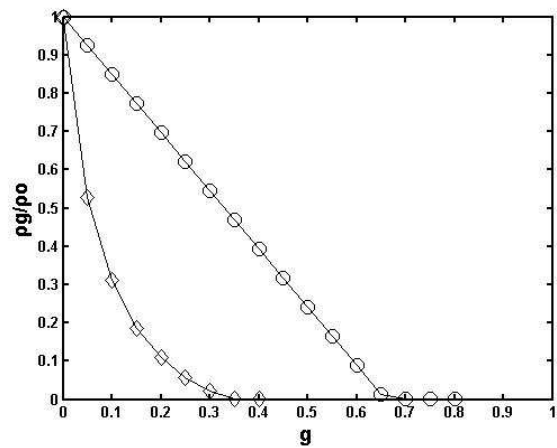}
  \centerline{(b)}\\
\caption{\label{F1}The relation of $\rho_g/\rho_0$ and density $g$
on the WS model with $N=10^4$, $\langle k \rangle=8$, $P=1$,
$\lambda=0.25$ (a) circles represent the random immunization and
diamonds represent the random ``network monitors", (b) circles
represent the targeted immunization and diamonds represent the
targeted ``network monitors".}
\end{figure}

We also conduct numerical simulation to test effectiveness of the
random immunization strategy. Besides random immunization, we also
test the effectiveness of the targeted immunization on the WS
network. The result can be seen in Fig.\ref{F1}. In Fig.\ref{F1}
(a), the data of circles represent the effectiveness of the random
immunization. It can be observed that nearly $0.8$ computer nodes
should be immunized in order to prevent the epidemic propagation. In
Fig.\ref{F1} (b), the data of circles represent the effectiveness of
the targeted immunization on the WS network. We should still
immunize nearly the fraction of $0.65$ nodes to prevent virus
propagation, which means targeted immunization is still not
efficient.

Consequently, if we want to efficiently prevent computer virus
propagation on homogeneous networks, we should install and update
antivirus programs for each computer nodes on the network. Then the
propagation behavior can be described by the SIS model. Relevant
studies indicated that random immunization and targeted immunization
are effective in this case \cite{2}.However, it can produce large
cost and work to install and update antivirus programs. In order to
avoid this problem, we introduce a new mechanism that can incur
relatively small cost. We call this new mechanism ``network
monitors". Specifically, we introduce two type ``network monitors".
One is the random ``network monitors" and the other is the targeted
``network monitors". The implementation of the random ``network
monitors" is as follows: we randomly select some nodes on the
network as ``network monitors". These nodes play two roles. On the
one hand, they are immunized nodes. In other words, we always update
new version antivirus programs and fire walls in ``network
monitors". So they cannot be infected by viruses. On the other hand,
``network monitors" can monitor their neighboring nodes. If they
find their neighbors infected by computer viruses, they can ``help"
infected neighbors recover to the susceptible nodes. In analysis, we
assume that each susceptible node is infected with rate $\lambda$
and the infective nodes can recover only by the help of ``network
monitors".

Now, we analyze the efficiency of random ``network monitors". Here,
we use $g$ to represent the density of ``network monitors". Then we
have the following mean-field equation for the random ``network
monitors":
\begin{equation}
\frac{\mathrm{d}\rho(t)}{\mathrm{d}t}=\lambda (1-g)\langle k \rangle
\rho(t)[1-\rho(t)]-f(\langle k \rangle ,g)\rho(t)
\end{equation}
In Eq.\,(5), $f(\langle k \rangle ,g)=$min$(1,\langle k \rangle g)$
is the probability that a given infected node connects with
``network monitors". After imposing the stationary condition
$\mathrm{d}\rho(t)/\mathrm{d}t=0$, we can obtain the following
quadratic equation of $\rho(t)$ :
\begin{equation}
\lambda(1-g)\langle k \rangle \rho(t)^2+[f(\langle k
\rangle,g)-\lambda(1-g)\langle k \rangle]\rho(t)=0
\end{equation}
$\rho(t)=0$ is the solution of Eq.\,(6). If we want Eq.\,(6) to only
have this trivial solution, we need $\Delta=0$. Then we can get the
critical density:\\
\begin{displaymath} g_c=
\begin{cases}
\lambda/(\lambda+1) & f(\langle k \rangle,g_c)=\langle k \rangle
g_c\\
1-1/(\lambda\langle k \rangle) & f(\langle k \rangle,g_c)=1
\end{cases}
\end{displaymath}
So we can conclude that our random ``network monitors" can
successfully suppress the epidemic outbreaks.

We also introduce the targeted ``network monitors". In this
mechanism, we choose those computer nodes with the highest degree as
``network monitors". In order to test the effectiveness of the
targeted ``network monitors", we perform numerical simulations.
Further, we also use numerical simulation to test the random
``network monitors".

Our simulations are as follows. In random ``network monitors", we
randomly choose some node to be infective individual and randomly
select nodes with density $g$ as ``network monitors". Then, if an
infective node connects to susceptible nodes, it infects each node
with the rate $\lambda$. As well, if an infective node is connected
to one or more ``network monitors", the infective node recovers to
the susceptible node. Finally, when the density of infective nodes
does not change, our experiment ends. In the simulation of the
targeted ``network monitors", the only difference from the random
``network monitors" is that we initially choose nodes with the
highest degree as ``network monitors". We perform simulations with
$1000$ different starting configurations and with at least $10$
different realizations of the network. Fig. \ref{F1} shows the
result of our numerical simulations.

In Fig.\ref{F1} (a), the data of diamonds represent the random
``network monitors". We observe that the critical density value $g$
is near $0.5$ for the random ``network monitors". This confirms our
analysis. Since we can get the critical density of random ``network
monitors" when $\langle k \rangle=8$ and $\lambda=0.25$ in analysis,
which is $g_c=1-1/(\lambda \langle k \rangle)=0.5$. In Fig.\ref{F1}
(b), diamonds represent the data of the targeted ``network
monitors". we observe that the targeted ``network monitors" is still
effective on the WS model.

\section{Eradicating epidemics on SF networks}
\label{sec:3} In this section, we focus on heterogeneous networks
especially on SF networks (we do not consider degree correlations
here). Firstly, we test whether the random ``network monitors" is
effective or not on SF networks. We write the mean-field rate
equation as:
\begin{equation}
\frac{\mathrm{d}\rho_k(t)}{\mathrm{d}t}=\lambda(1-g)
k(1-\rho_k(t))\Theta(t)-f(k,g)\rho_k(t)
\end{equation}
Here, $g$ represents the density of ``network monitors" and the
function $f(k, g)$ can be written as
$f(k,g)=$min$(1,k\\\sum\limits_{k'}k'p(k')g_{k'}/\langle k
\rangle)$, which is the probability that an infected node with k
links connects with some ``network monitor". In addition,
$\rho_k(t)$ represents the density of infected nodes that have
degree $k$ and we define the quantity
$\Theta(t)=\sum\limits_kkp_k\rho_k(t)/\sum\limits_ssp_s$ as the
probability of the susceptible node pointing to an infected node at
time $t$. In Eq.\,(7), we observe that if we can keep $f(k,
g)\equiv1$, then the random ``network monitors" is most efficient.
It is apparent that if above condition is fulfilled, the effect of
random ``network monitors" is equivalent to the effect of random
immunization strategy in the SIS model. From Ref. \cite{2} we know
that this immunization strategy is not particularly effective on SF
networks in the SIS model. As a consequence, random ``network
monitors" is also not effective on SF networks.

Next, we investigate whether the targeted ``network monitors" is
effective or not on SF networks. After applying the targeted
``network monitors", we obtain the following mean-field equation:
\begin{equation}
\frac{\mathrm{d}\rho_k(t)}{\mathrm{d}t}=\lambda
k(1-\rho_k(t)-g_k)\Theta(t)-f(k,g)\rho_k(t)
\end{equation}
In Eq.\,(8), $g_k$ represents the density of ``network monitors"
with degree $k$ and the function $f(k, g)$ can be written as
$f(k,g)=$min$(1,k\sum\limits_{k'}k'p(k')g_{k'}/\langle k \rangle)$.
Here we assume that after applying the targeted ``network monitors",
the probability that a node with $k$ links is healthy is
$(1-\rho_k(t)-g_k)$. In the stationary state
($\mathrm{d}\rho_k(t)/\mathrm{d}t=0$) we obtain Eq.\,(9) from
Eq.\,(8):
\begin{equation}
\lambda k(1-\rho_k-g_k)\Theta=f(k,g)\rho_k
\end{equation}
From Eq.\,(9) we can get the solution of $\rho_k(t)$ in the
stationary state: $\rho_k=((1-g_k)\lambda k\Theta)/(f(k,g)+\lambda
k\Theta)$. Now inserting this result into the definition of
$\Theta$, we can obtain Eq.\,(10):
\begin{equation}
\Theta=\frac{1}{\langle k
\rangle}\sum\limits_kkp(k)\frac{(1-g_k)\lambda
k\Theta}{f(k,g)+\lambda k\Theta}
\end{equation}
Here, we define an auxiliary function
\begin{displaymath}
F(\Theta)=\frac{1}{\langle k
\rangle}\sum\limits_kkp(k)\frac{(1-g_k)\lambda
k\Theta}{f(k,g)+\lambda k\Theta}-\Theta
\end{displaymath}
It is apparent that $F(0)=0$. Since $\frac{1}{\langle k
\rangle}\sum\limits_kkp(k)\frac{(1-g_k)\lambda
k\Theta}{f(k,g)+\lambda k\Theta}\leq\frac{1}{\langle k
\rangle}\sum\limits_kkp(k)\frac{(1-g_k)\lambda k\Theta}{\lambda
k\Theta}<1$, we have $F(1)<0$. If $F'(0)<0$, then $F(\Theta)<0$ for
some $\Theta\in(0,1]$. But if $F'(0)\geq0$, there is a
$\Theta_1\in(0,1)$ that satisfies $F(\Theta_1)>0$. Then by the
intermediate theorem, there is a $\Theta\in(\Theta_1,1)$ satisfies
$F(\Theta)=0$. Consequently, if there is a $\Theta\in(0,1)$ that
satisfies $F(\Theta)=0$, $F'(0)\geq0$. It is equivalent to say that
if there is a non-zero solution of Eq.\,(10), then:
\begin{equation}
\frac{\mathrm{d}}{\mathrm{d}\Theta}(\frac{1}{\langle k
\rangle}\sum\limits_kkp(k)\frac{(1-g_k)\lambda
k\Theta}{f(k,g)+\lambda k \Theta})\mid_{\Theta=0}\geq 1
\end{equation}
So we can get Eq.\,(12):
\begin{equation}
\frac{\lambda}{\langle k
\rangle}\sum_k\frac{k^2(1-g_k)p(k)}{f(k,g_c)}=1
\end{equation}
Here $g_c$ represents critical density of ``network monitors". We
define a new quantity $k_t$. When $k>k_t$ the density of ``network
monitors" $g_k=1$ and when $k\leq k_t$ the density of ``network
monitors" $g_k=0$. So we rewrite Eq.\,(12) as follows:
\begin{equation}
\frac{\lambda}{\langle k
\rangle}\sum\limits_{k=0}^{k_t}\frac{k^2p(k)}{f(k,g_c)}=1
\end{equation}

From the definition of $k_t$ we can get
$g_c=\sum\limits_{k>k_t}p(k)\approx1-\int\nolimits_0^{k_t}p(k)\mathrm{d}k$.
Then we can obtain the expression of $k_t$ as a function of $g_c$.
Finally, we insert this expression into Eq.\,(13) to get the
critical density $g_c$.

In order to support our analysis, we perform numerical simulations
on the BA network \cite{17} with the network size $N=10^4$, $m_0=7$,
$m=4$ and $\lambda=0.25$. Initially, we infect only one node.

\begin{figure}
\centering
\resizebox{0.75\columnwidth}{!}{
  \includegraphics[width=8.3cm]{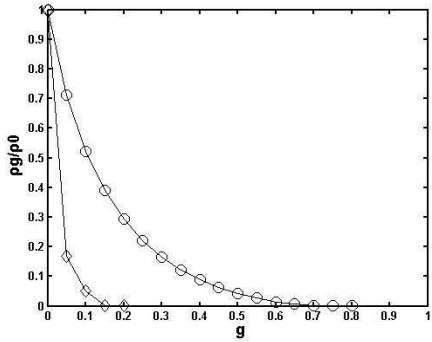}
  }
\caption{\label{F2} The relation of $\rho_g/\rho_0$ and density $g$
on the BA model with $N=10^4$, $m_0=7$, $m=4$, $\lambda=0.25$.
Circles represent the strategy of random ``network monitors" and
diamonds represent the strategy of  targeted ``network monitors".}
\end{figure}

Fig. \ref{F2} shows the relation between reduced density of infected
nodes $\rho_g/\rho_0$ in the end and the density of ``network
monitors" $g$. In Fig. \ref{F2}, circles represent the data of the
random ``network monitors". We observe that the random ``network
monitors" is not particularly effective. We should immunize large
proportion of nodes as monitors to eradicate infection. On the
contrary, we find that the targeted ``network monitors" is
effective. This fact can be observed in Fig. \ref{F2}, in which the
critical density of ``network monitors" $g_c\approx0.15$.

\section{Discussion and conclusions}
\label{sec:4} The main conclusion can be summarized as follows:\\
\qquad $1.$ If we can only install and update the antivirus programs
with the fraction $\alpha$ ($0<\alpha<<1$) computer nodes on the
network, the random immunization and the targeted immunization are
not efficient to prevent computer viruses spreading on the
homogeneous
networks.\\
\qquad $2.$ We introduce a policy to protect the computer networks,
namely, ``network monitors". Specifically, we choose some computer
nodes on the network as the ``network monitors". They play two
roles. On the one hand, They are immune to computer viruses. On the
other hand, ``network monitors" can monitor their neighboring nodes'
behaviors. If they find their neighbors infected by computer
viruses, they can eliminate computer viruses from infected
neighbors. Based on different approach to select nodes as the
``network monitors", we divide the ``network monitors" into two
types, one is the random ``network monitors" and the other is the
targeted ``network monitors". Through analysis, we find that the
random ``network monitors" and the targeted ``network monitors" are
effective on WS networks. Besides, the targeted ``network monitors"
is effective on SF networks. Numerical simulations confirm our
analysis.

In fact, our new strategy is similar with the contact tracing
\cite{18} but we explicitly set monitors on the network. If
possible, we can initially install antivirus programs in each
computer node on the network and then only update antivirus programs
in ``network monitors" . So the recovering probability of the
infective computer node is generally not zero, since old version
antivirus programs can ``kill" ordinary viruses. Then the number of
``network monitors" chosen to eradicate computer viruses is smaller
than our analysis. It demonstrates that the strategy of the
``network monitors" is practical and efficient that we need only
update small fraction computers' antivirus programs and fire walls
so that the security of the whole computer network can be kept.

The author thanks Robert Ellis, Hemanshu Kaul and Michael Pelsmajer
for meaningful comments.

%
%

\end{document}